\documentstyle[11pt,newpasp,twoside,epsf]{article}
\def\edcomment#1{\iffalse\marginpar{\raggedright\sl#1\/}\else\relax\fi}
\reversemarginpar

\begin{document}
\title{Radiative Transfer in Spiral Galaxies}
 \author{Simone Bianchi}
\affil{Istituto di Radioastronomia/CNR - Sez. di Firenze, 
L.go E. Fermi 5, 50125, Firenze, Italy}

\begin{abstract}
The internal dust extinction in spiral galaxies can affect our understanding 
of their structure and morphology, as well as our perception of the distant
universe in the background. The intrinsic properties of the stellar and dust
components can be studied by comparing  the observed appearance of a dusty
spiral in the optical with a radiative transfer model. The absorbed starlight
is re-radiated by dust in the FIR/sub-mm. Thus, observations at long 
wavelengths allow to put further constraints on the properties of the dust
distribution. I present our Monte Carlo simulations for dust extinction and
emission in galactic disks, discuss the basic ingredients and limitations and
compare them with other solutions for the radiative transfer available in the
literature. I review the current picture of dusty disks that emerges from
radiative transfer studies.
\end{abstract}

\section{Introduction}

For many extragalactic astronomers, dust is a nuisance: intrinsic
luminosities and morphology, star formation rates, face-on corrections
and several other quantities characterizing a spiral galaxy may be
affected by dust extinction, requiring a correction to retrieve them 
from observed data. Furthermore, dusty disks or dust ejected into the
intergalactic medium could hamper our view of the distant universe. 
However, dust is also interesting for its own sake. Dust is the 
repository of a good fraction of the metals in a spiral, and 
any model for galaxy evolution and star formation history should
be constrained by the present amount of grains. Dust plays an
active role in the interstellar medium, e.g. as a site of molecule
formation or as a source of heating photoelectrons, and it may play
a significant role in the intergalactic medium as well, if dust grains
can escape the potential well of a galaxy.

Before the advent of infrared observations, the amount of dust in 
a spiral galaxy was traditionally quantified by it extinction
effects\footnote{For reviews, see Calzetti (2001) or the 
introductions in the radiative transfer papers listed in the 
next section (e.g.\ Witt, Thronson \& Capuano 1992; 
Byun, Freeman \& Kylafis 1994)}. Starting from Holmberg (1958),
statistical studies of the surface brightness in samples of galaxies 
offering different inclinations to the observers were taken as a proof 
for a substantial transparency of galactic disks (at least when not seen 
edge-on).  However such studies, besides being possibly affected by subtle 
selection effects, also depend on the geometric model assumed for the
relative distribution of dust and stars: in a provocative paper, 
Disney, Davies \& Phillipps (1989) shook the traditional view by
showing how the same galactic samples were compatible with models 
completely opaque to radiation.

Therefore, to asses the amount of dust and its effects in a spiral
galaxy disk it is necessary to solve the radiative transfer problem for 
geometries as close as possible to those inferred from observations.

\section{Radiative Transfer vs Geometry}

For light traveling in a dusty medium along a path of length $s$ and 
direction $k$, the change in the intensity $I_\lambda$ of radiation is 
described by the well known radiative transfer equation (Chandrasekhar 1960),
\begin{equation}
\frac{d I_\lambda }{d s} = -\kappa_\lambda I_\lambda +j_\lambda
+\kappa_\lambda \frac{\omega_\lambda}{4\pi} \int I_\lambda(k') 
\Phi_\lambda(k,k') d\Omega.
\end{equation}
The first term on the right hand side of the Eq.~1
represents the light that is removed from the beam, either because of 
absorption by or scattering off dust grains (both effects included in 
the {\em extinction} coefficient $\kappa_\lambda$). The opacity of a medium
is in general characterized by the optical depth $\tau_\lambda$, the 
integral of $\kappa_\lambda$ along the whole radiation path.
The other two terms in Eq.~1 are positive contributions 
to $I_\lambda$, due to direct sources of radiation (the emission coefficient 
$j_\lambda$) or to light coming from different directions $k'$ that is 
scattered into the direction $k$. The last term depends on the dust phase 
function $\Phi_\lambda$ (stating the directionality of the scattering) and 
the albedo $\omega_\lambda$ (the fraction of extinguished light that is 
scattered).

It is relatively easy to solve the radiative transfer equation omitting the 
scattering term (at least numerically). Unfortunately, this is not justified 
a priori in the UV-Optical-NIR regime (where $j_\lambda$ is mainly due to 
starlight), since a good fraction of the light impinging on each grain
is scattered (for Milky Way dust $\omega_V=0.6$; Gordon, Calzetti \& Witt 
1997; Gordon, these proceedings)\footnote{Conversely, solutions are
simpler for FIR/submm radiation, where emission is mostly due to dust: for 
particles much smaller than the wavelength of light scattering is negligible
({\em Rayleigh} scattering; van de Hulst 1957).}

To ease the solution of the radiative transfer equation inclusive of 
scattering, several authors have adopted a plane parallel geometry to 
describe the galactic disk (see Baes \& Dejonghe 2001 for a review and
a comparison between different methods).
Two popular plane parallel model are the {\em Slab}, in which stars and
dust are homogeneously mixed, and the {\em Sandwich}, in which the dust
distribution is thinner than the stellar distribution.

However, the surface brightness distribution in spirals is better 
described by exponential disks (Freeman 1970; Wainscoat, Freeman \&
Hyland 1989).
Two methods have been used to solve the radiative transfer problem
in such geometry. The first is the one proposed by Kylafis \& Bahcall (1987;
Byun et al.\ 1994; Xilouris et al.\ 1999). Eq.~(1) can be
solved separately for $I_0$, the contribution due to directly emitted
photons, $I_1$, the contribution due to photons that have scattered
only once, $I_2$, the contribution due to photons that have scattered
only twice, etc. The total intensity is the summation of all terms. In
the  Kylafis \& Bahcall method $I_0$ and $I_1$ are computed exactly,
while the contribution due to scattering events of order $n\ge2$ is
approximated assuming that $I_n/I_{n-1} = I_1/I_0$.

\begin{figure}
\plotfiddle{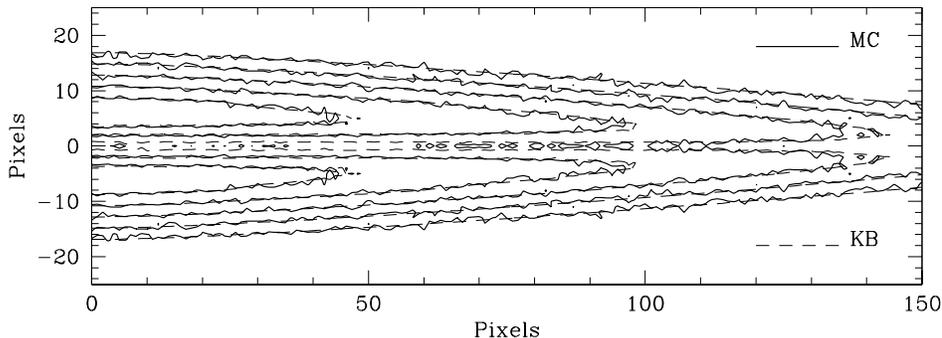}{4.5cm}{0}{65}{65}{-210}{-110}
\caption{Radiative transfer simulation of an edge-on dusty disk.
Solid contours are from a Monte Carlo model, dashed contours
refer to the Kylafis \& Bahcall (1987) approach (Misiriotis \& Bianchi
2001).}
\end{figure}

The second approach is the Monte Carlo method (Bianchi, Ferrara 
\& Giovanardi 1996; De Jong 1996; Wood \& Jones 1997; Baes et al.\ 2003).
The life of each photon is followed as it moves through dust and
its fate is derived in a probabilistic way by tossing random numbers.
The position where the photon is emitted is drawn from the adopted
distribution of sources; a path direction is assigned, in general
assuming isotropic emission. On a given path of optical depth 
$\tau$, the probability for a photon to avoid absorption/scattering 
is $e^{-\tau}$; thus a {\em probable} optical depth along the path can be 
derived. If such optical depth is larger than the total optical depth
along the photon traveling direction, the photon escape the dust
distribution and can be recorded by an {\em observer}, otherwise it encounters 
a grain inside the dust distribution. Here the photon can be either 
absorbed or scattered along 
another direction. If the photon is scattered (a fraction $\omega$ of all 
photons impinging on a grain are scattered) the phase function $\Phi$ is used 
as the probability distribution for the angle between the old and new 
traveling direction. The process is then repeated on the new path. 
Several cycles (photons) are needed to produce high S/N results (integrated 
photometry or maps as in Fig.~1). This basic scheme can be optimized to 
reduce the number of cycles needed to achieve the same S/N (Gordon et al.\ 
2001; Baes et al.\ 2003). The Monte Carlo procedure is
quite general and has been used to study radiative transfer in other
geometries, like spheroidals which may be appropriate for starburst
galaxies and ellipticals (Witt et al.\ 1992; Gordon et al.\ 1997;
Baes \& Dejonghe 2002) or young stars in their birth 
environment (Whitney, these proceedings).

Several other works have adopted radiative transfer recipes to deal with 
dust extinction, especially when modeling the spectral evolution in 
galaxies (see, for example, Charlot \& Fall 2000; Granato et al.\ 2000).
I have limited the review here only to models focused on the radiative 
transfer problem.

\section{Radiative Transfer in a Homogeneous Medium}

In radiative transfer models, the stellar disk is usually reproduced
by a double exponential,
\begin{equation}
\rho=\rho_0 \exp \left(-\frac{r}{\alpha_\star} -\frac{|z|}{\beta_\star}\right),
\end{equation}
where $\alpha_\star$ and $\beta_\star$ are the radial and vertical
scalelengths of the distribution. Observations in the Milky Way and in
other spirals suggest $\alpha_\star/\beta_\star \approx 10-15$ (Bahcall
\& Soneira 1980; De Grijs \& Van Der Kruit 1996) with trends of smaller
$\beta_\star$'s (Mihalas \& Binney 1981) and larger $\alpha_\star$'s
(de Jong 1996) for bluer starlight (although the radial color gradients 
can also be explained with extinction; Peletier et al.\ 1995). In
analogy with stars, Eq.~2 is also adopted for the dust distribution, with
independent scalelengths $\alpha_d$ and $\beta_d$. A common assumption
is that $\alpha_d=\alpha_\star$ and $\beta_d \approx 0.5 \beta_d$ (the
latter needed to reproduce the extinction lanes observed in edge-on
galaxies; Fig.~1). I will call this the standard model and parametrise the 
amount of dust in the disk by assigning  a value to $\tau_V$, the V-band 
optical depth along the face-on direction through the disk plane.

In Bianchi et al.\ (1996) we have described the behavior of standard
models with different $\tau_V$, extending the study to a larger set of
parameters in Ferrara et al.\ (1999). A spheroidal distribution was also
included to simulate the bulge in late types. For our first work we have
used the Draine \& Lee (1984) model to derive the dust properties 
($\omega_\lambda$, $\Phi_\lambda$ and extinction law; the last needed to
scale the opacity of models at different $\lambda$'s to $\tau_V$).
Later, we have preferred an empirical approach, with extinction laws,
albedos and phase functions derived from observations of Milky Way 
dust (Gordon et al.\ 1997; these proceedings).

\begin{figure}
\plotfiddle{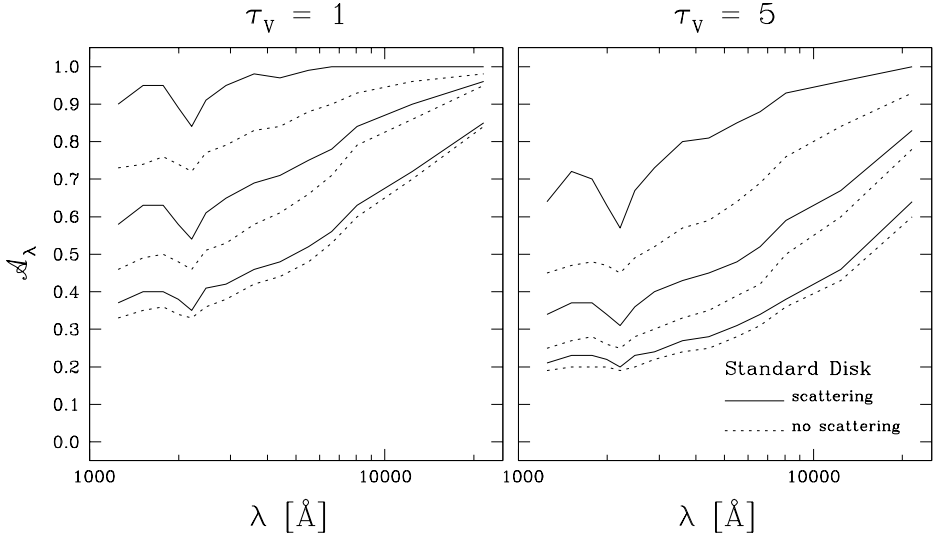}{7.0cm}{0}{70}{70}{-210}{-165}
\caption{Attenuation curves for a standard disk (model {\tt SXX\_ME04}, B/T=0 
in Ferrara et al.\ 1999) with face-on optical depths $\tau_{V}=1$ (left
panel) and 5 (right panel). Solid lines show models with scattering, dotted 
lines models without scattering. For each model three inclinations are 
presented ($i=30\deg$, 80$\deg$ and 90$\deg$ from top to bottom).}
\end{figure}

Once images of simulated galaxies are produced, the variations of observable 
quantities (e.g. total magnitudes, surface brightnesses, colour gradients)
can be analyzed as a function of the inclination and opacity of the disk
(Byun et al.\ 1994). An important quantity that can be computed from the models
is the {\em attenuation} $\cal{A}_\lambda$, i.e.\ the ratio between the 
flux observed from a galaxy and the flux that would be observed in 
the absence of dust. To illustrate the importance of scattering and its 
effects, we show in Fig.~2 the behavior of $\cal{A}_\lambda$ for two standard 
disks of different opacities ($\tau_V=1$ and 5) and seen at different 
inclinations and wavelengths (solid lines). When scattering is not included 
(dotted lines) the attenuation is clearly overestimated. The effect is more 
important for disks seen face-on: due to the higher optical depth of any path 
along the plane of the disk, light is more easily scattered above the plane 
than in the edge-on direction. 

The trend of $\cal{A}_\lambda$ with wavelength reflects the trend in the 
Milky Way extinction law, which has been used for the models of Fig.~2. 
The main feature in $\cal{A}_\lambda$ is the footprint of the 2175\AA\
bump, whose strength and width depend on the geometry, dust opacity and 
inclination. The feature's $\lambda$ dependence is in
general smoother than one would expect for a dust screen of the same 
opacity in front of the galaxy, since in a galaxy model the observed flux 
includes light coming from different regions and having suffered different 
absorption and scattering.  For the same reason, 
further smoothing is possible when considering clumpy dust distributions 
(next section). However, it is usually believed that radiative transfer 
alone cannot smooth the bump out completely (Vijh, Witt \& Gordon 2003; 
Clayton, these proceedings). Thus, the featureless {\em Calzetti law} 
(Calzetti 2001) derived on starburst galaxies implies a different dust 
composition or size distribution in those objects (but see Granato et 
al.\ 2000 for a different view).

The attenuation is larger for models with larger $\tau_V$ (and for larger
inclinations) but it does not increase indefinitely with increasing opacity.
For high optical depths a saturation is reached when stars high above the 
plane of the galaxy, and thus unextinguished, become the dominant component 
of the galactic flux. For the same reason, the colour of a galaxy does not 
become indefinitely redder for increasing $\tau_V$, as it would be for a
screen of dust in front of a light source (Witt et al.\ 1992).

In Bianchi et al.\ (1996) the study of radiative transfer was extended to
polarization due to scattering off spherical grains. While in some peculiar
cases it is possible to reproduce the linear polarization pattern parallel
to the galactic plane observed in the Milky Way and in most spirals 
(Whittet, these proceedings), for normal dust size distributions the
linear polarization due to scattering is perpendicular to the disk. 
Polarization due to aligned grains is needed (Wood \& Jones 1997), but so
far a proper treatment of radiative transfer including dicroic extinction 
and scattering by elongated grains has only been applied to
circumstellar environments (Whitney \& Wolff 2002).

\section{Radiative Transfer in a Clumpy Medium}

A smooth exponential model is only a first approximation for galactic
disks, which are observed to have a complex clumpy structure. Clumping
can be easily incorporated in Monte Carlo simulations by dividing the
three dimensional space in cells and assigning to each of them a clumpy
or smooth status.  In Bianchi \& al.\ (2000c) and Misiriotis \& Bianchi 
(2002) we modelled the dust distribution as a two phase medium,
including a smooth exponential disk and a distribution of clouds.
We assumed that the clumpy dust medium is associated with the molecular
gas component of a spiral disk and we allowed for different fractions of 
dust mass in clumps (25, 50 and 75\%) analogous to the fraction of
gas mass in the molecular component in late-type galaxies; the
mass and dimension (size of each cell) of the clumps were derived from
Milky Way giant molecular clouds; the distribution of clumps
was taken to be a ring, as for HII in the Milky Way, or an exponential,
as for the molecular gas of most late type spyrals.

\begin{figure}
\plotfiddle{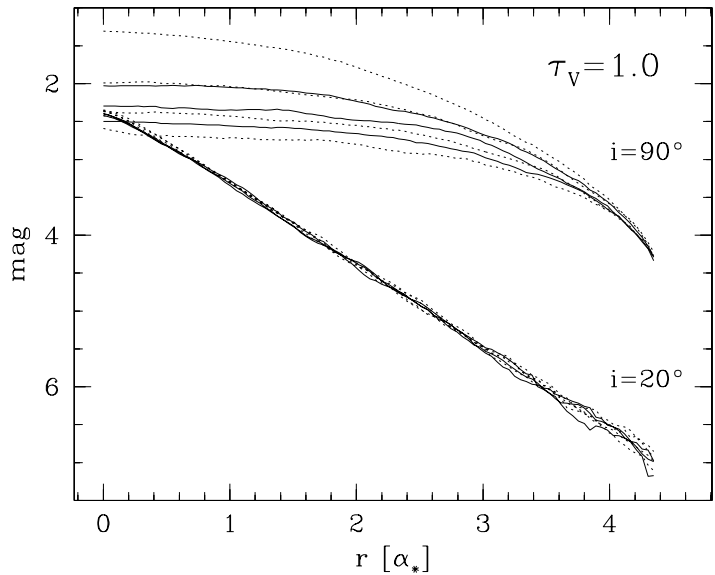}{7.5cm}{0}{55}{55}{-170}{-90}
\caption{V-band major axis profiles for a clumpy disk with dust 
mass equivalent to that of a $\tau_V$=1 homogeneous standard disk.
Clumps have the same exponential distribution as 
the diffuse dust.  Profiles are shown for inclinations $i$=20$\deg$ 
(nearly face-on) and 90$\deg$ (edge-on). Solid lines refer to models with 
25\%, 50\%, 75\% of the dust mass in clumps, from bottom to top, 
respectively. For comparison, we also show profiles for homogeneous 
disks (dotted lines) with $\tau_V$= 0.25, 0.5, 0.75, 1 
(from top to bottom).
}
\end{figure}

As expected, models in which the distribution of dust is clumpy have a
higher transparency than models in which the same amount of dust is in
a homogeneous disk. However, for the parameter range we have explored
the effects of clumping are moderate, with changes in attenuation laws 
and in the amount of absorbed energy of less than 30\% in the V-band. 
In Fig.~3 we
show the radial surface brightness profiles of clumpy models with the
same amount of dust as a $\tau_V=1$ homogeneous model. The distribution
of clumps is exponential. Because of the low optical depth, in the face-on 
case there is hardly any difference between homogeneous and clumpy models.
Instead, profiles for edge-on disks with clumping look brighter than
the corresponding homogeneous one: for instance, the model with 75\% of
the dust in clumps has the same profile as a more transparent, $\tau_V=0.5$,
model. This is interesting, since comparisons between models and observations 
are more easily done at high inclinations, where the effects of dust are 
maximized and the geometric components of a galaxy can be easily separated
(next section): if clumping is not taken into account, the amount of dust 
in a galaxy may be underestimated.

Unfortunately, clumping depends on the adopted parameterization. In the
galaxy models of Kuchinski et al.\ (1998), for example, homogeneous and 
clumpy models of edge-on galaxies are very similar.
They follow the Witt \& Gordon (2000) formalism for clumping, with 
constant cloud filling factors and density contrast with respect to the 
smooth medium. For the parameters they adopted (based on a comparison with 
the cloud mass spectra in the Galactic ISM) most of the dust mass 
($\approx$95\%) is in clumps. As a result, any line of sight along the 
galactic plane will intersect a larger number of clumps than in our
models, making little difference between the clumpy and homogeneous case.

I recall here that these general considerations are valid when clumping 
affects only the extinguishing medium. If we allow for a fraction of the 
starlight to be emitted inside clouds (as it is the case for young stars), 
models with clumping could be less transparent than a homogeneous one, 
depending on the amount of embedded radiation (Bianchi et al.\ 2000c).

\section{Radiative Transfer vs Observations}

A direct comparison between radiative transfer models for homogeneous
dusty disks and observations of edge-on galaxies was done by Xilouris et
al.\ (1999). Using the Kylafis \& Bahcall approach, they fitted 
the surface brightness distribution of seven galaxies observed in the 
optical/NIR and obtained the parameters for the dust (disk) and stellar 
(disk + bulge) components. They derived a mean optical depth for the
sample, $< \tau_V >  \approx 0.6$, with the opacities in the various 
optical/NIR bands showing an extinction law similar to that in the Milky Way. 
Moderate-to-low opacities are compatible with those obtained with a
variety of methods for the diffuse dust component (Calzetti 2001): 
the dust disks, thus, would be almost transparent when seen face-on 
and unable to produce a significant obscuration of the distant universe 
(Alton et al.\ 2001). As for the relative dust/star geometry, Xilouris et
al.\ found that: $<\beta_d/\beta_\star> \approx 0.5$, necessary
to explain the dark extinction lane; $<\alpha_d/\alpha_\star> 
\approx 1.4$, the dust disk thus being more extended than the stellar disk.

Tests on mock data have shown how disk dishomogeneities cannot 
severely bias a fit made with homogeneous distributions: the neglect of 
spiral structure only affects the derived parameters by less then few 
percent (Misiriotis et al.\ 2000); clumping can produce a moderate dust 
underestimation by $\la$30\% in mass (somewhat less than predicted 
on the sole basis of the major axis profile; Misiriotis \& Bianchi 2002) 
or even less for the clumpy disks of Kuchinski et al.\ (1998). 

Other studies support extended dust disks: a large dust disk is invoked
by Peletier et al.\ (1995) to explain the change of B-K colour gradients
with the inclination in a sample of 37 galaxies; the FIR emission
observed in the Milky Way by the COBE satellite can be explained with
an exponential dust disk with $\alpha_d/\alpha_\star$=1.5 (Davies et al.\
1997); a dust spatial distribution less concentrated than the stellar 
is suggested by coadded IRAS 100$\mu$m images of spiral galaxies
(Nelson, Zaritsky \& Cutri 1998).

Even more striking were the results of Alton et al.\ (1998b). They
observed a sample of seven nearby resolved spirals at 200 $\mu$m
with the ISOPHOT camera on board of the ISO satellite. Large amounts of 
cold dust ($T\approx 20K$) could be detected, which were previously missed 
by the satellite IRAS operating at wavelengths below the peak of dust emission.
The scalelength in 200$\mu$m maps was larger than that observed by IRAS 
at 100$\mu$m, as expected for diffuse 
dust heated by the Interstellar Radiation Field (colder dust at larger
distances from the centre, where the ISRF is less intense). However, the 
ISO emission was also found to be as extended as or broader than the 
B-band emission, implying a dust intrinsic scalelength 
larger than the stellar ($\alpha_d/\alpha_\star > 1$). How much
larger?

\section{Models of FIR Emission}

The increasing availability of FIR data allows a more direct study of
the dust content of a galaxy than extinction measures. To be able to
compare radiative transfer models with observation at longer wavelengths, 
in Bianchi, Davies \& Alton (2000a) we extended our simulations to include 
emission from dust heated by the ISRF. In particular, we were interested in 
veryfing if the large emission scalelengths observed by Alton et al.\ (1998b) 
were compatible with the values of $\alpha_d$ derived by Xilouris et al.\
(1999) for the intrinsic dust distribution.

The first step in a model of dust emission is computing the local heating 
ISRF or, correspondingly, the amount of energy that is absorbed by dust 
grains as a function of position inside the galaxy. This is straightforward 
in a Monte Carlo simulation where, for each light-grain interaction, the
amount of energy absorbed and the position of the event is known. Once a
Spectral Energy Distribution (SED) for starlight is adopted, several 
monochromatic simulations are run and a final map of the total energy 
absorbed from starlight can be produced.

More complex is deriving the dust temperature resulting from the heating. 
Sophisticated approaches will consider both thermal equilibrium emission 
from large grains and stochastic heating of small grains and PAHs
(Popescu et al.\ 2000; Misselt et al.\ 2001). Since we were interested 
mainly in emission at $\lambda > 100\mu$m, we only considered the former.
As stellar radiation is absorbed by all kind of grains, we used the
dust model of Desert, Boulanger \& Puget (1990) to derive the amount of
energy that goes into non equilibrium emission (about 20-30\% of the 
infrared dust output) and subtracted it from the map of total energy
absorbed. Then, the dust temperature was computed using an emissivity
law derived from observations of the  Milky Way FIR emission (Bianchi,
Davies \& Alton 1999). The same emissivity also tells us that FIR
opacities are very low, at least for the diffuse medium we are dealing
with. Thus, the optically thin regime can be considered: maps of
emission can be produced by simply integrating the emission coefficient 
(a modified blackbody) along the chosen line of sight (i.e. for the
FIR/submm, only $j_\nu$ is left in the right hand side of Eq.~1).

In Bianchi et al.\ (2000a) we produced simulations for the the spiral galaxy
NGC~6946, whose stellar SED is well covered by observations from the UV
to the NIR. For each geometry and $\tau_V$ we chose, we carried out the 
radiative transfer for several wavelengths and we normalized the
simulations at each $\lambda$ to produce the same output in radiation as
observed. That is, we derived {\em a posteriori} the intrinsic, dust-free,
emission that would result in the observed, dust-extinguished, flux.
As a result, the intrinsic dust-free stellar SED depends on the
adopted geometry and $\tau_V$. Dust in NGC~6946 is observed to emit 
$\approx$30\% of the galaxy bolometric luminosity. In this, NGC~6946 is
very similar to other late type spirals in the nearby universe (Popescu 
\& Tuffs 2002). In a realistic simulation, then, dust needs to be able 
to absorb 30\% of the stellar radiation. Such comparison between stellar 
and dust energy outputs is usually referred to as {\em energy balance}.

I show here the results for an optically thin ($\tau_V$=0.5) and
thick ($\tau_V$=5) standard ($\alpha_d/\alpha_\star$=1.0) model and an 
optically thick ($\tau_V$=5) extended ($\alpha_d/\alpha_\star$=1.5) model. 
All models have homogeneous exponential stellar and dust distributions
with $\beta_d/\beta_\star$=0.5 and it is assumed that the relative 
stellar/dust geometry do not change with the wavelength of the starlight. The
results are not significantly different if $\beta_\star$ is allowed to 
decrease with $\lambda$ or if a bulge is included (both cases having
a slightly larger efficiency in absorbing starlight for the same
$\tau_V$ because more stellar emission is inside the dust distribution; 
Bianchi 1999).

The temperatures distributions are very similar from model to model, with 
values similar to those derived in other galaxies and gradients that are 
compatible with those derived in the Milky Way (Fig.~4). A look at the
SED for dust emission (Fig.~5) shows that the optically thin standard
model does not absorb enough stellar radiation (only 5\%) to match the 
observed emission. For dust to be able to absorb between 25 and 40\% of
the total bolometric luminosity, a standard model with optical depth 
$5 \la \tau_V \la 10$ is needed. For a disk as in Xilouris et al.\
(1999) slightly more radiation is absorbed than in a standard disk
with the same $\tau_V$: the extended $\tau_V$=5 model fits reasonably
well the observed data by absorbing 30\% of starlight (of which 70\% is 
emitted at thermal equilibrium). The optically thin models inferred from
observations, instead, do not have the necessary FIR energy output.
The need for high optical depths for 
the energy balance in spirals is confirmed by other authors
(Evans 1992; Trewhella 1998; Radovich, Kaahanp{\" a}{\" a} \& Lemke
2001; in the the first two works the neglect of scattering may result
in an overestimate of the absorption; the last work is a Monte Carlo
radiative transfer simulation very similar to our own). 

Popescu et al.\ (2000) and Misiriotis et al.\ (2001) modelled the 
dust emission for 5 of the 7 galaxies whose parameters where derived by
Xilouris et al.\ (1999). Since they lack data in the UV, they provide
the fitted stellar distributions with an extra, UV emitting, disk thinner
than the dust distribution (i.e.\ $\beta_\star^{UV}/\beta_s < 1$).
The Star Formation Rate in the UV disk is derived by allowing the model
to fit the FIR
emission in the two IRAS bands at 60$\mu$m and 100$\mu$m. The fitted
values for the SFR are consistent with SFR normally observed in
spirals. Without such UV emitting disk, the stellar and dust
distributions of Xilouris et al.\ (1999) are not able to emit enough
FIR flux. Unfortunately only for two objects (namely, NGC~891 and
NGC~5907) they have datapoints at $\lambda$'s beyond IRAS data. When
these submm observations are considered, the model is not able to predict
the corresponding flux, unless a supplementary disk of dust is
included. If the new disk is as thin as the UV disk, it would be
undetected in fits of edge-on surface brightness profiles, being
obscured by the more vertically extended dust disk. Neglecting 
the new dust disk, the fits of Xilouris et al.\ (1999) underestimate
the total dust mass by a factor between 2 and 4.

Although the sample analised is limited, a common trend seems to emerge: 
the dust disks implied by the surface brightness distributions of edge-on 
galaxies do not provide enough absorption to explain the FIR emission. 
The discrepancy cannot be easily explained with clumping, at least with
the clumping discussed in Sect.~4. A different parametrization for clumping 
of dust and sources may be needed. For instance, the thin extra UV and dust 
disks introduced by Popescu et al.\ (2000) and Misiriotis et al.\ (2001) 
may hide a distribution of sources still embedded in giant clouds. While
clumping of dust does not modify significantly the FIR SED (Misselt et
al.\ 2001), preferential emission within dust clouds could increase the 
efficiency of a dust disk in absorbing (and emitting) radiation.

\begin{figure}
\plotfiddle{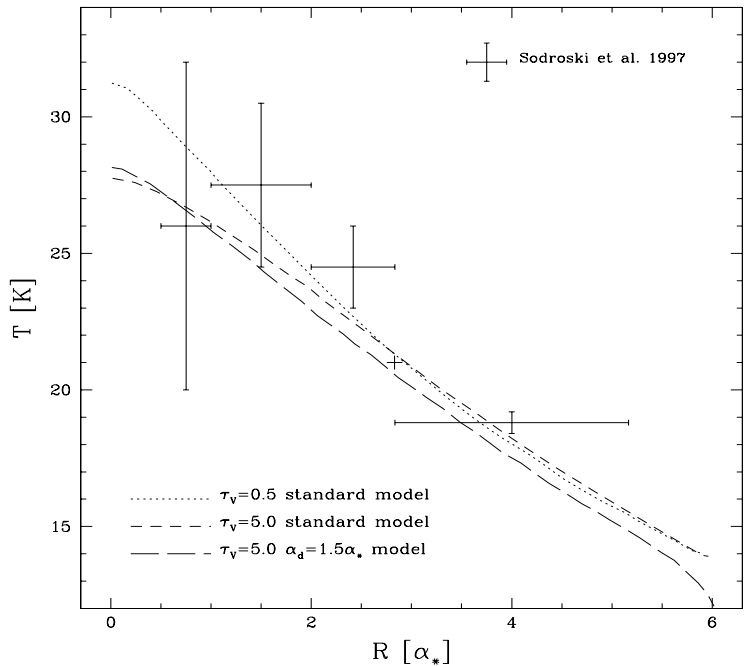}{9.0cm}{0}{55}{50}{-160}{-80}
\caption{Temperature gradient along the galactic plane for three 
different models (see text). Data points show the radial gradient of
temperature in the Galaxy (derived from Sodroski et al.\ 1997).
The cross marks the temperature of 21K at the Sun distance from
the galactic centre (Bianchi et al.\ 1999).}
\end{figure}

What about the spatial distribution of FIR emission?
Because of the temperature gradient, emission at longer 
wavelength is characterized by a larger emission scalelength, approaching 
the intrinsic radial dust scalelength $\alpha_d$ as $\lambda$ increases. 
While the extended disk model for NGC~6946 has a $100\mu$m 
scalelength similar
to that observed in IRAS data, the broad $200\mu$m emission observed by 
Alton et al.\ (1998b) cannot be explained. Models that could reproduce
the $200\mu$m emission ({\em only}) require too broad distributions that
are ruled out by observations (Bianchi et al.\ 2000a). Recently, concern
has been raised on the results of Alton et al.\ (1998b): the mapping procedure
with which the galaxy sample was observed is known to artificially broaden 
the detected emission when scanning over a bright source (Popescu et al.\ 
2002). However, the broadening should be less important across the scan 
direction, where Alton et al.\ measured essentially the same scalelengths 
as those along the scan.

\begin{figure}
\plotfiddle{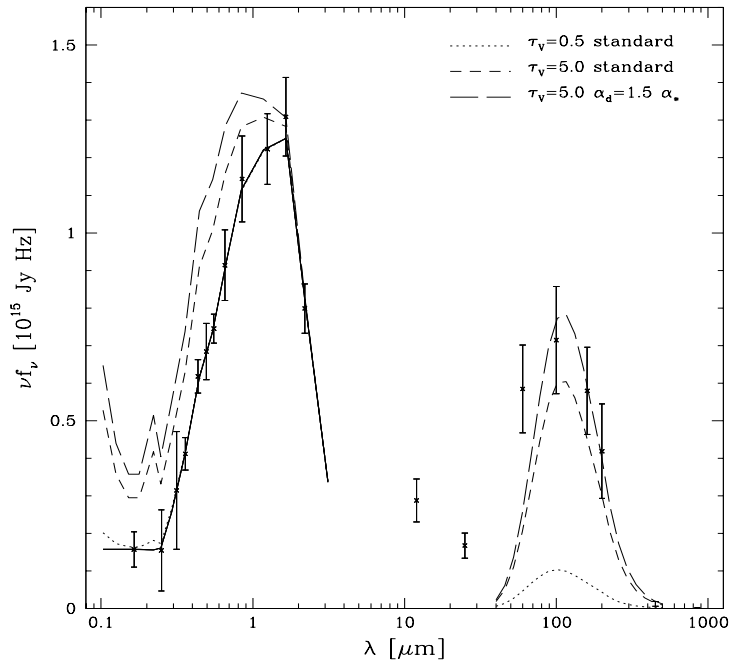}{9.0cm}{0}{55}{50}{-160}{-80}
\caption{Observed and modelled SED of NGC~6946. The models are the same 
as in Fig.~4. The solid line is the stellar SED derived from the 
observations (data points), which is the common UV-optical-NIR output
for all models. The intrinsic, dust-free, stellar SED (shown for each
single model in the figure) is derived {\em a posteriori} from the radiative 
transfer simulations. The spike visible in the intrinsic stellar SEDs is due 
to the 2175\AA\ extinction feature, which is present in the MW extinction law 
assumed for dust but absent in the observed SED. The SED of dust
emission is shown for each model in the FIR.
More details in Bianchi et al.\ (2000).
}
\end{figure}

\section{The Future}

Despite the progress in modelling, a clear picture of dust extinction and 
emission in {\em real} 
spiral galaxies is not yet available. A better understanding of the 
problem will be possible when a detailed wavelength coverage of a
galaxy SED is available, for a large number of nearby objects well
resolved at each $\lambda$. Data around the peak of dust emission are
obviously needed. A move in this direction is the SIRTF Nearby Galaxy
Survey (SINGS; Kennicutt et al.\ 2003). Future SIRTF observations of a 
sample of 75 nearby objects will be integrated with other data from 
archives and supplementary programs, providing an unprecedented view
on both dust emission and stellar emission. The latter will extend down
to the UV, which is very important to understand if FIR emission in local
universe spirals is a good tracer of recent star formation (Walterbos \& 
Greenawalt 1996). The radiative transfer models of Popescu et al.\ (2000) 
and Misiriotis et al.\  (2001) propend for this hypothesis, while in our
work dust preferentially absorbs radiation from old stars emitting in
the optical/NIR (Bianchi et al.\ 2000a).

Good resolution in observations of dust emission is an advantage, 
especially to define the influence of clumping. Available submm instruments 
like the camera SCUBA can provide an adequate resolution but they are not 
sensitive enough to detect emission from diffuse regions, thus limiting the 
study to dense regions in face-on galaxies or to the inner disk in edge-on 
objects: for instance in the face-on NGC~6946 only emission associated with 
dense molecular gas could be detected (Bianchi et al.\ 2000b), while cold 
dust could be observed up to \twothirds\ of the optical disk in the edge-on 
NGC~891 (Alton et al.\ 1998a). More sensitive and high resolution
observations in the submm and in the FIR will be possible in the nearby
future, from a variety of planned ground-based or satellite-born
instruments and facilities (SCUBA2, APEX, SOFIA, Herschel, ALMA).

Together with the improvements in the observing capabilities, more
refined radiative transfer models have to be developed, also gaining 
insights on the dust/stellar geometry by analyzing other observables 
affected by extinction, like polarization and observed kinematics 
(for the last issue, see the recent radiative transfer studies by 
Matthews \& Wood 2001 and Baes et al.\ 2002, 2003).

\vspace{0.2cm}
\acknowledgments I deeply thank my advisors and collaborators in 
this projects: A. Ferrara, C. Giovanardi, J.~I. Davies, P.~B. Alton
and A. Misiriotis.

\end{document}